# The auditory Tuning of a Keyboard

A possible well tempered optimum

Johan Broekaert ; broekaert.devriendt(at)telenet.be
http://users.telenet.be/broekaert-devriendt/Index.html

KEYWORDS : Well, Temperament, Tuning, Auditory, Keyboard, Optimum, Bach

ABSTRACT : An optimal auditory tunable well (circular) temperament is determined. A temperament that is applicable in practice is derived from this optimum. No other historical temperament fits as well, with this optimum. A brief comparison of temperaments is worked out.

## [1]    The Auditory Tuning

The initial octave intonation for the auditory tuning of a keyboard, relies mainly on the tuning of beating rates of fifths and major thirds, within a chromatic scale ranging from F3 to F4 (Calvet, 2020).

Interval beatings are the product of interfering harmonics (sinusoidal sound waves) of the sounds of musical notes. Those harmonics are often called "partials", especially if their frequencies deviate from integer multiples of the first partial frequency. Some partials of notes of an interval might have almost equal frequencies. For example : the third partial of a note, and the second partial of the fifth on that note will have equal or almost equal frequencies. Two simultaneous sinusoidal sound waves of almost equal frequency sound like one wave at median frequency, amplitude modulated at a frequency equal to the frequency difference : "the beating".

Kellner (1977) has evaluated a temperament based on calculated interval beating rates. See tables 1 and 2, calculating interval beating rates of fifths and major thirds within the F3–F4 scale.

| | | | |
|---|---|---|---|
| $q_F = 2C4 - 3F3$ | $q_C = 4G3 - 3C4$ | $q_G = 2D4 - 3G3$ | $q_D = 4A3 - 3D4$ |
| $q_A = 2E4 - 3A3$ | $q_E = 4B3 - 3E4$ | $q_B = 4F\#3 - 3B3$ | $q_{F\#} = 2C\#4 - 3F\#3$ |
| $q_{C\#} = 4G\#3 - 3C\#4$ | $q_{G\#} = 2Eb4 - 3G\#3$ | $q_{Eb} = 4Bb3 - 3Eb4$ | $q_{Bb} = 4F3 - 3Bb3$ |

Table 1: Calculated beating rate of fifths on notes within the F3–F4 scale

| | | | |
|---|---|---|---|
| $p_F = 4A3 - 5F3$ | $p_C = 4E4 - 5C4$ | $p_G = 4B3 - 5G3$ | $p_D = 8F\#3 - 5D4$ |
| $p_A = 4C\#4 - 5A3$ | $p_E = 8G\#3 - 5E4$ | $p_B = 4Eb4 - 5B3$ | $p_{F\#} = 4Bb3 - 5F\#3$ |
| $p_{C\#} = 8F3 - 5C\#4$ | $p_{G\#} = 4C4 - 5G\#3$ | $p_{Eb} = 8G3 - 5Eb4$ | $p_{Bb} = 4D4 - 5Bb3$ |

Table 2: Calculated beating rate of major thirds on notes within the F3–F4 scale

## [2]    Well Tempering (circulating temperaments)

Well tempering strives for good purity of the diatonic C–major, and acceptable purity for all other tonalities. A possible C–major impurity measure can be the following sum of squared beating rates :

$$\sum (q_{Note}^2 + p_{Note}^2) = q_F^2 + q_C^2 + q_G^2 + q_D^2 + q_A^2 + q_E^2 \quad + \quad p_F^2 + p_C^2 + p_G^2$$

The minimum of this sum is obtained by evaluating the sum in function of the notes, followed by setting the partial derivatives to the notes equal to zero. This leads to a soluble set of linear equations. This innovating musical analysis procedure, was suggested by Prof. E. Amiot.

The beating rates within this thus obtained temperament are low but quite dissimilar, and the dissimilarity might be a musical drawback and complicates tuning. A lower rate of the fastest beatings can be desired, even if this is slightly at the expense of the slowest beatings.

## [3] The Optimal Beatings Equality

It is musically better, and also easier for auditory tuning, to estimate the equality of slow beatings, rather than their minimum sum. Indeed, tuning for a minimum impurity requires that impurities should be measured, summated and compared to a known minimum ; this is hard to implement.

A good equality corresponds to a small deviation from an average value "M". Beatings are normally negative on fifths (too small), and positive on thirds (too large). Normally, the absolute average "M", therefore is:

$$M = \frac{-q_F - q_C - q_G - q_D - q_A - q_E + p_F + p_C + p_G}{9}$$

Taking the signs into account, the deviations from the mean beating are :

Fifths : $\Delta_{Fi;Note} = -q_{Note} - M$      Major Thirds : $\Delta_{T;Note} = p_{Note} - M$

The sum of the squares of these deviations becomes:

$$\sum \Delta^2_{Fi\ and\ T;Note} = \Delta^2_{Fi;F} + \Delta^2_{Fi;C} + \Delta^2_{Fi;G} + \Delta^2_{Fi;D} + \Delta^2_{Fi;A} + \Delta^2_{Fi;E} + \Delta^2_{T;F} + \Delta^2_{T;C} + \Delta^2_{T;G}$$

The elaboration of this sum, expressed in function of note pitches, gives :

$$81 \times \sum \Delta^2_{Fi\ and\ T;Note} = 2718 F_3^2 + 2934 C_4^2 + 3726 G_3^2 + 1044 D_4^2 + 3240 A_3^2 + 2124 E_4^2 + 2592 B_3^2$$
$$- 1116 F_3 C_4 - 216 F_3 G_3 + 36 F_3 D_4 - 3132 F_3 A_3 + 180 F_3 E_4$$
$$- 2376 C_4 G_3 + 72 C_4 D_4 + 216 C_4 A_3 - 2880 C_4 E_4$$
$$- 864 G_3 D_4 + 324 G_3 A_3 + 540 G_3 E_4 - 3240 G_3 B_3$$
$$- 1998 D_4 A_3 - 90 D_4 E_4 \quad - 1242 A_3 E_4 \quad - 1944 E_4 B_3$$

The partial derivatives of this sum equated to zero, lead after simplification of coefficients, to the set of equations, table 3 :

The five remaining notes, F#3, C#4, G#3, Eb4, Bb3, and their beatings, are obtained by setting an equal beating for the six involved fifths; the major thirds do not have to be optimized, those have rapid beatings anyhow :

|  | F3 | C4 | G3 | D4 | E4 | B3 | = | A3 |
|---|---|---|---|---|---|---|---|---|
| ∂/∂F3 : | 151 | −31 | −6 | 1 | 5 | 0 | = | 87 |
| ∂/∂C4 : | −31 | 163 | −66 | 2 | −80 | 0 | = | −6 |
| ∂/∂G3 : | −2 | −22 | 69 | −8 | 5 | −30 | = | −3 |
| ∂/∂D4 : | 2 | 4 | −48 | 116 | −5 | 0 | = | 111 |
| ∂/∂E4 : | 10 | −160 | 30 | −5 | 236 | −108 | = | 69 |
| ∂/∂B3 : | 0 | 0 | −5 | 0 | −3 | 8 | = | 0 |

Table 3: Calculation of diatonic notes for C – major

$$q_{Note} = q_B = q_{F\#} = q_{C\#} = q_{G\#} = q_{Eb} = q_{Bb}$$

The collection of solutions gives the scale, table 4:

|  | F3 | F#3 | G3 | G#3 | A3 | Bb3 | B3 | C4 | C#4 | D4 | Eb4 | E4 | F4 |
|---|---|---|---|---|---|---|---|---|---|---|---|---|---|
| $f_{Note}$ | 175.67 | 184.73 | 196.60 | 207.98 | 220.00 | 234.14 | 246.22 | 262.75 | 277.22 | 293.96 | 312.10 | 328.93 | 351.34 |
| $q_{Note}$ | −1.52 | 0.26 | −1.89 | 0.26 | −2.15 | 0.26 | 0.26 | −1.83 | 0.26 | −1.87 | 0.26 | −1.89 | −3.04 |
| $p_{Note}$ | 1.65 | 12.91 | 1.89 | 11.07 | 8.90 | 5.13 | 17.30 | 1.97 | 19.23 | 8.06 | 12.30 | 19.23 | 3.30 |

Table 4: scale with optimal beating rate equality of the of diatonic fifths and major thirds within C major

The average beating rate for fifths and major thirds between F3 and F4, is 1.85, with only minor deviations from this value (≤ 0.33). The six remaining fifths are equal and slightly enlarged.

## [4] A possible "Optimal" Equal Beating Rate, in Practice

[4.1] Jobin (2005) has developed a temperament holding five equal fifths (if calculated in cents), and two pure major thirds, based on an interpretation of scrolls on a figure on a score of "Das wohltemperirte Clavier" of J. S Bach. He reports the necessity for minor auditory corrections when tuning this temperament. Based on his paper and the above section [3], one could wonder : why not strive for five fifths and two thirds with almost equal beating rate ?

When calculating the optimal equality of beatings of the major thirds C–E, G–B and the five fifths on which those are built, according to the method in sections [2] and [3], one unexpectedly encounters a perfect equality. And also the major third F–A can still be set equal : the F note was not yet involved in calculations.

It is therefore easier to calculate these notes, by directly assuming the equalities. A system of eight equations with seven unknowns is obtained. This system is redundant by chance or luck, and it suffices to solve the first seven equations:

$$q_{Note} = q_C = q_G = q_D = q_A = q_E = -p_C = -p_F \quad = \quad -p_G = -p_{Note}$$

These perfect equalities provide for additional ease of tuning, and lead to a dependent F–C fifth, which fortunately has sufficient quality.

[4.2] Further notes.

F#, C# and G# are determined by perfect fifths on B, F# and C#, this facilitates tuning, and tonalities with sharps are favored in terms of purity over those with flats.

$$q_{Note} = 0 = q_B = q_{F\#} = q_{C\#}$$

Only the notes E*b* and B*b* need still to be tuned, most easily by even division of the remaining minor impurity over the fifths on A*b* (G#), E*b* and B*b* :

$$q_{Note} = q_{Ab} = q_{Eb} = q_{Bb}$$

[4.3] The collection of solutions gives the following scale :

|  | F3 | F#3 | G3 | G#3 | A3 | B*b*3 | B3 | C4 | C#4 | D4 | E*b*4 | E4 | F4 |
|---|---|---|---|---|---|---|---|---|---|---|---|---|---|
| $f_{Noot}$ | 175.61 | 184.71 | 196.64 | 207.80 | 220.00 | 234.02 | 246.28 | 262.83 | 277.07 | 293.98 | 311.90 | 329.03 | 351.22 |
| $q_{Noot}$ | − 1.17 | 0.00 | − 1.95 | 0.39 | − 1.95 | 0.39 | 0.00 | − 1.95 | 0.00 | − 1.95 | 0.39 | − 1.95 | − 2.34 |
| $p_{Noot}$ | 1.95 | 12.51 | 1.95 | 12.32 | 8.27 | 5.84 | 16.17 | 1.95 | 19.54 | 7.79 | 13.62 | 17.28 | 3.89 |

Table 5: scale with quasi optimal beating rate equality for diatonic fifths and major thirds

Figure 1 illustrates "unconventionally" the fifths and thirds impurities in beatings/sec, within a chromatic scale on C4 ; the course within this figure is fairly irregular.

Figure 2 illustrates a "conventional" representation of the fifths and major thirds impurities in cents, and a rather normal course can be perceived, very comparable to those of famous well temperaments. All fifths are purer than the purest thirds, and the impurity of slightly augmented fifths on A*b* (G#), E*b*, B*b*, turns out to be insignificant.

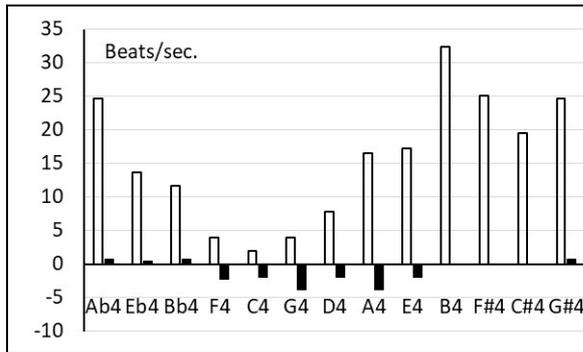
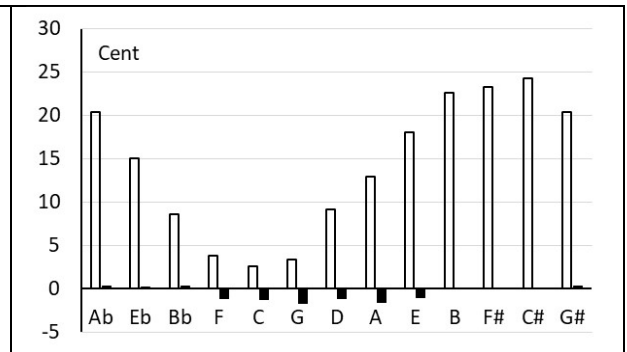

Figure 1: Impurities of fifths (in black) and thirds

Figure 2: Impurities of fifths (in black) and thirds

figure 3 "unconventionally" illustrates the fifths and thirds impurities in beatings/sec, within a chromatic scale on F3, the scale usually used for auditory tuning. The notes on the figure are given in an "unconventional" inverted sequence of fifths. The displayed course is remarkable and regular for the natural notes of C–major.

Figure 4 illustrates in a similar way, the impurities expressed in cents. It has exactly the same course as fig. 2, but with an inversion and "phase shift" of the sequence of notes.

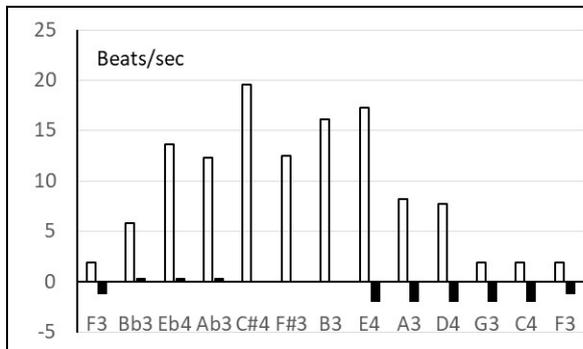
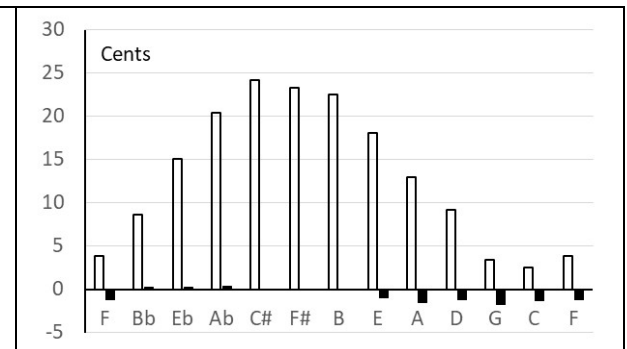

Figure 3: Impurities of fifths (in black) and thirds

Figure 4: Impurities of fifths (in black) and thirds

## [5]     Comparison of temperaments

Characteristics of any temperament are unambiguously defined by either: the note pitches, or the semi–tones, fifths, fourths or sevenths (=11 semi–tones) characteristics. Fifths characteristics are most commonly applied. This way, comparison of temperaments can be based on a measure of the fifths differences in cents, possibly by applying the formula below :

$$RMS_{Difference} = \sqrt{\frac{\sum_{F,C,G,D,A,E,B,F\#,C\#,G\#,Eb,Bb}(fifth_{Note;1} - fifth_{Note;2})^2}{12}}$$

The $RMS_{Difference}$ of the temperament section [4] and the optimal one section [3] amounts to only 0.42 cents. This difference amounts to 0.98 cents for the "classical" cent calculated Vallotti, an 1.30 cents for the "equally beating" Vallotti temperament ($q_{F,C,G,D,A,E}$ = – 2.27 beats/sec., for A=440). All other temperaments have higher deviations

## [6]     "Das wohltemperirte Clavier"

Well tempering becomes often related to "Das wohltemperirte Clavier" (1722; 1740–42), a masterpiece of J. S. Bach.

It was discussed, sometimes indirectly, by Kirnberger (1771), in the letters of Kirnberger to Forkel (Kelletat, 1960, 1981, 1982), and by Forkel (1802), all arguing in favor of some kind of well temperament.

It was also discussed by Marpurg (1776), arguing in favor of the equal temperament (12TET). The assumption of Marpurg was copied by a countless number of authors in a countless number of publications, over two centuries, and even today.

A probable first doubt concerning the application of the 12TET by Bach was published by Bosanquet H. (1876), and a breakthrough about those doubts probably came about through Kelletat (1960). Since then many assumptions are published about possible "Bach–temperaments". To cite the most famous only, we can (chronologically) think of Kelletat (1966), Kellner (1977), Billeter (1979. 2008), Sparschuh (1999), Zapf (2001), Jobin (2005), Lehman (2005), Lindley (2006), Amiot (2008), and many others for sure are missing in this summary list (Calvet and Lehman for instance, publish longer lists).

Figure 5 illustrates the fifths impurities course in beatings/sec., of the temperament defined in section [4], in a same sequence as on figures 3 and 4. This is also the sequence in which they can be assumed on a figure of a score by J. S. Bach: an "inverted" fifths sequence, within the F3–F4 range, – this is the notes range also used for tuning–. The scrolled part of figure 5 is copied from Amiot (2008).

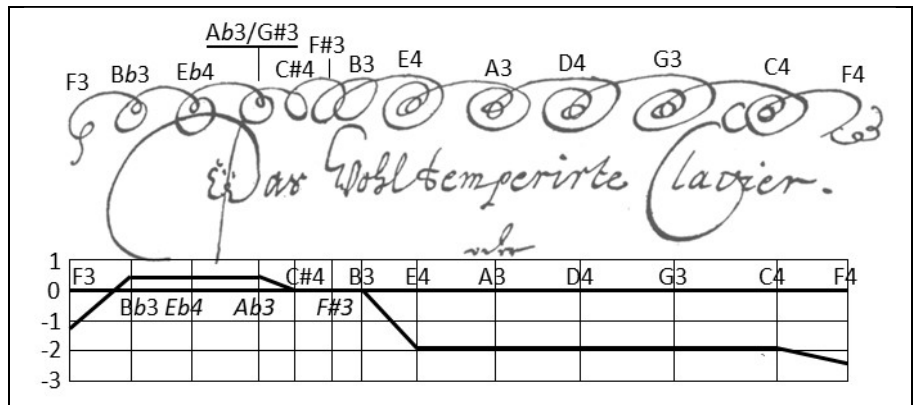

Figure 5: course of impurities of fifths, in beatings/sec., drawn in front of the same course on a figure on a score by J. S. Bach

One can observe a striking parallelism between the note marks on the scrolls and the notes beatings course, on the graph below the scrolls.

ACKNOWLEGMENTS

My most sincere thanks to Prof. E. Amiot and A. Calvet, auditory tuner. It would not at all have been possible to write the present analysis without their competent professional input and support. Thanks also to E. Jobin, my source of inspiration for section [4].

Thanks to my daughter Hilde. She suggested to investigate on "What does a musician (tuner) want"? (cfr. the significant difference for impurity evaluations between sections [2] and [3]).